# Ideological and Policy Origins of the Internet, 1957-1969

Presented to 29[th] TPRC, October 28, 2001, Alexandria, Virginia
Andrew L. Russell
Department of History, University of Colorado at Boulder
andrew.russell@colorado.edu

> [T]hese guys at MIT and BBN… We'd all gotten really excited about interactive computing, and we had a kind of little religion growing here about how this was going to be totally different than batch processing… I had this little picture in my mind of how we were going to get people and computers really thinking together. [ARPA director Jack] Ruina was thinking in terms of command and control, and it didn't take really very much to see how this would work.  I wanted to get all this done.  So I had the notion [that] 'command and control essentially depends on interactive computing so the military really needs this.'  I was one of the few people who, I think, had this positive feeling toward the military. It wasn't just to fund our stuff, but they really needed it and they were good guys. So I set out to build this program.

> - J. C. R. Licklider[1]

## Introduction

Value-sensitive design is a forward-looking enterprise, aiming to integrate human values into the design of technologies.[2]   Its insights also prove useful as a way to illuminate the values already embedded in technologies.  This paper, as a historical analysis of the foundational values of ARPA's Information Processing Techniques Office (IPTO) and the ARPANET, is an attempt to build a bridge between the intellectual and policy history of the Cold War and the value-sensitive design of cyberspace.

---

[1] J. C. R. Licklider, quoted in John A. N. Lee and Robert Rosin, "The Project MAC Interviews," *IEEE Annals of the History of Computing*, Vol. 14, No. 2, 1992, 16-17. BBN is Bolt, Beranek and Newman, a Cambridge, MA computer consulting firm.  ARPA is the Advanced Research Projects Agency, later renamed DARPA (Defense Advanced Research Projects Agency).

[2] Batya Friedman defines value-sensitive design as "an approach to the design of technology that accounts for human values in a principled and comprehensive manner throughout the design process," focused especially on human welfare and justice. See Batya Friedman, ed. *Human Values and the Design of Computer Technology* (New York: Cambridge University Press and CSLI, Stanford University, 1997); Deborah G. Johnson and Helen Nissenbaum, eds., *Computers, Ethics and Social Values* (Englewood Cliffs, NJ: Prentice-Hall, 1995); and Peter G. Neumann, "Psychological implications of computer software development and use: Zen and the art of computing," in Domenico Ferrari, Mario Bolognani & Joseph Goguen, eds., *Theory and Practice of Software Technology* (New York: North-Holland, 1983), 221-232.

By analyzing U. S. science policy and the development of ARPA and IPTO between 1957 and 1969, I argue that the ideological framework of the science policy consensus shaped the institutions and relationships between the government and computing that led to the creation of IPTO, and eventually to the ARPANET. During this time period, the values of this broader consensus created space for individual computer scientists to incorporate their own values into the creation of IPTO and the design of cyberspace.

Consensus—among science policymakers and among academic and government computer researchers—was a dominant founding principle of IPTO research between 1957 and 1969. How this "consensus" operated within the creation and activities of IPTO influenced how it would evolve as a directing principle for the design of computer networks and cyberspace. Three areas are particularly helpful for understanding how "consensus" values directed the creation of the ARPANET. First, the ideology of the post-World War II science policy consensus, finely focused after the launch of *Sputnik*, demonstrates how cooperation between universities, industry, and the government guided scientific research in the first half of the Cold War. Second, the hiring in 1961 of J.C.R. Licklider to direct the IPTO enabled him to coordinate military and academic computer research around his concept of "interactive computing." Third, the arrival in 1966 of Larry Roberts at IPTO and the launching of the ARPANET highlights how ARPA could leverage the science policy consensus to achieve its advanced research mission.

This examination of the values of the stakeholders in IPTO research can help us to understand the ways that these values shaped the creation of the ARPANET. The influence of the values expressed by these actors was decisive. They believed that government had an obligation to support a broad base of scientific research to promote both the public good and the national defense; that IPTO-sponsored computing research would accomplish both military and scientific objectives; and that IPTO could leverage its power within this consensus to create a network to share resources and unite researchers over geographical distance.

## Science policy "consensus," *Sputnik*, and ARPA

Historians note a self-conscious "consensus" among both American policymakers and scientists in the late 1940s and early 1950s.[3] In the wake of Nazi Germany and in the face of the Cold War Soviet "menace," Americans fashioned and clung to what historian Arthur Schlesinger has called a "vital center" in order to resist the threat of totalitarianism from the right and the left, and, ultimately, to secure the survival of American liberal democracy into the 1950s and beyond.[4] Policymakers and historians identify a science policy consensus operating among scientists and policymakers during this same period in the first 20 years of the Cold War. Among the tenets of this broad consensus was that the American victory in World War II owed much to the vitality of basic science research, especially advances in physics that led to microwave radar, proximity fuses, solid-fuel rockets, and the atomic bomb. The indisputable lesson was that "national security depended upon research in pure science."[5]  The premise that the federal government should completely pull away from its support of science research during peacetime was never a serious scientific or political option.[6]  The historian J. Merton England stressed this very point by introducing his seminal *A Patron for Pure Science: The National Science Foundation's Formative Years, 1945-1957* with the words of Montaigne: "Men only debate and question of the branch, not of the tree."[7]

---

[3] The existence of such a consensus does not deny the existence of widespread resistance to dominant social trends. For less benign aspects of these dominant social trends, see especially Jessica Wang, *American Science in an Age of Anxiety: Scientists, Anticommunism, and the Cold War* (Chapel Hill: The University of North Carolina Press, 1999) and Elaine Tyler May, *Homeward Bound: American Families in the Cold War Era* (New York: Basic Books, 1988).
[4] Arthur M. Schlesinger, Jr., *The Vital Center: The Politics of Freedom* (New Brunswick, NJ: Transaction Publishers, 1949. Reprint, 1998), Robert Griffith, "Dwight D. Eisenhower and the Corporate Commonwealth," *The American Historical Review*, Volume 87, Issue 1, Supplement (Feb., 1982).

Historian Bruce Smith notes that the central features of the science policy consensus reveal Vannevar Bush's seminal influence on its character.  Five ideas particularly shaped science policy: basic research would drive the system; commercialization and the creation of new markets should follow "almost automatically" government investments in basic research; the control of science through legislation was of less importance than promoting the growth of scientific research; technology would play a crucial role in foreign policy, to maintain military superiority and as an enabler of free trade; and the ideological support for the consensus came with the widespread faith in the progress of science and understanding that the "growth of federal research served everybody's interests—universities, government agencies, industry, [and] congressional committees."[8]

---

[5] Daniel J. Kevles, "Cold War and Hot Physics: Science, Security, and the American State, 1945-56," *Historical Studies in the Physical Sciences* 20, no. 2 (1990): 240.

[6] Daniel J. Kevles, "The National Science Foundation and the Debate over Postwar Research Policy, 1942-1955," *Isis* 68, no. 241 (1977): 4-26.  The legislative debates that occurred during the creation of the NSF were concerned more with the details of how the agency would run: whether the military or civilians should oversee this agency, whether it should include medical research, the degree to which federal support should be concentrated or geographically distributed, and how the agency should handle wartime patents.

[7] J. Merton England, *A Patron for Pure Science: The National Science Foundation's Formative Years, 1945-1957* (Washington, D.C.: National Science Foundation, 1983), 1.

[8] Bruce L. R. Smith, *U. S. Science Policy Since World War II* (Washington, DC: The Brookings Institution, 1990), 39.  David Hart, in his excellent book *Forged Consensus*, demonstrates how Bush's influence on this consensus, while significant, is overemphasized in most accounts of the subject. Hart demonstrates how figures in American politics and economics "forged" this consensus throughout the course of the twentieth century, from Herbert Hoover's "associationalism" to the New Deal liberalism of Karl Compton and Thurman Arnold. David Hart, *Forged*

The launch and orbit of *Sputnik I* on October 4, 1957 created psychological and scientific shock waves in the science policy community, as well as throughout American politics and the American public. *Sputnik* crystallized political perceptions of the need—both psychological and technical, especially in light of *Sputnik*'s implications for intercontinental ballistic missiles (ICBMs)—for fundamental breakthroughs in American science. As a result, federal spending on science research increased dramatically in the decade after *Sputnik*, prompting scientists and historians to label these years a "golden age" of American science.[9]

Almost immediately after *Sputnik*, President Dwight Eisenhower created ARPA as the agency responsible for coordinating advanced, high-risk research that could lead to technological breakthroughs for the military at the height of the Cold War. Founded in early 1958, ARPA was an institutional manifestation of a combination of three factors: Eisenhower's management style, the *Sputnik*-inspired technological crisis, and the science policy consensus. The creation of ARPA, as an agency outside of the three services yet responsible for coordinating advanced R&D for military applications, was a direct manifestation of Eisenhower's desire to "strike a balance among the competing claims of the services," especially in the areas of space and ballistic missile R&D.[10] While prominent scientists such as James Killian pressed during the 1950s for the

*Consensus: Science, Technology, and Public Policy in the United States, 1921-1953* (Princeton, NJ: Princeton University Press, 1998).

[9] Between 1959 and development budget almost tripled, from under $5 billion to over $13 billion. See Bruce L. R. Smith, *American Science Policy since World War II* (Washington, DC: The Brookings Institution, 1990) and Richard Barber Associates, *The Advanced Research Projects Agency, 1958-1974* (Washington, DC: Barber Associates, 1975), I-4. According to Harvey Brooks, 1967 was high-water mark of the "golden age of academic and basic science of the early 1960s." Harvey Brooks, "Are Scientists Obsolete?", *Science* vol. 186, 8 November 1974, p. 501. MIT professor and Lincoln Lab researcher Wesley Clark remembered the 1960s as "kind of a heyday of government funding." See also Wesley Clark interview, May 3, 1990, in New York, NY, conducted by Judy O'Neill, from transcript, Charles Babbage Institute, University of Minnesota, Minneapolis.

[10] Robert Griffith, "Dwight D. Eisenhower and the Corporate Commonwealth," *The American Historical Review*, Volume 87, Issue 1, Supplement (Feb., 1982), 96.

Department of Defense (DoD) to reorganize R&D and to further support the basic research that would prevent the U.S. from "falling behind" the Soviets, *Sputnik* was the decisive catalyst for the creation of an agency like ARPA to coordinate advanced military R&D pushing the frontiers of science and technology.[11] As a 1975 ARPA-commissioned institutional history suggested, "Thus ARPA was to be spawned in an environment which equated basic science with military security."[12]

Existing histories of ARPA and IPTO explain the creation of ARPA as a product of the *Sputnik* scare, but they do not mention how the creation of ARPA was a logical outgrowth of the consensus, developed throughout the twentieth century, of dominant American policy positions on the role of the federal government in promoting science for the general welfare and for the national defense. "These ideas were shaped," reported the ARPA-commissioned history, "throughout the early 1950s, largely by a group of scientists who had come to fear that government either misused or misunderstood modern science and technology… When the bell rang—Sputnik—they were ready with a rather complete agenda and proceeded to act on it."[13]

**Computers for the national defense, 1945-1961**

One of the most vivid sites of the operation of this consensus can be seen in the support of computing research by the DoD throughout the Cold War. Since World War II, scientists in the military and in government recognized the relevance of computers to the national defense and explored ways that advanced computing could enhance national security. One example is the SAGE air defense system, developed during the 1950s by MIT scientists, funded by the Air Force, as part of an effort to use computers to work with people, in real time, to solve complex problems. SAGE's use of computers—in an immediate, intelligent, interactive manner—was a departure from the standard contemporary use for computers as batch-processing number-crunchers. While SAGE thus set a technological precedent for large, interactive computer systems, it also set the administrative precedent of a computer research project that was of high value both to the academic research community and to the military.[14]

---

[11] Smith, *American Science Policy*, 36-72.
[12] Barber Associates, *ARPA*, I-26.
[13] Barber Associates, *ARPA*, I-23.

Developments in Cold War politics and in the technical aspects of computing coincided to create an opportunity to advance the state of computing. In March 1961 President John F. Kennedy called on Congress to fund improved command and control systems that would be "more flexible, more selective, more deliberate, better protected, and under ultimate civilian authority at all times."[15] As it worked out, strong policy support for science research for the national defense occurred just as significant developments in computer hardware, timesharing, displays, and programming were changing the nature of computing itself.[16]

**J. C. R. Licklider and Man-Computer Symbiosis as a creation of the consensus**

The most significant product of the consensus for computer research came with the creation of IPTO. With the arrival of the Kennedy administration in 1961, ARPA

---

[14] Thomas P. Hughes, *Rescuing Prometheus* (New York: Pantheon Books, 1998).
[15] "Message from the President of the United States Relative to Recommendations Relating to Our Defense Budget," 87th Congress, 1st Session, 28 March 1961, House Documents, Document No. 123, 8, cited in Arthur L. Norberg and Judy E. O'Neill, *Transforming Computer Technology* (Baltimore: The Johns Hopkins University Press, 1996), 10.
[16] See Paul E. Ceruzzi, *A History of Modern Computing* (Cambridge, MA: The MIT Press, 1998).

Director Jack Ruina and Eugene Fubini[17] decided to centralize the DoD's research on computing technology into one program. Ruina asked Joseph C. R. Licklider, Professor of Psychology at MIT's Lincoln Lab, to lead the office. Licklider, known to Ruina as a "man of some distinction,"[18] had previous experience in military science through acoustic research for the Navy during World War II, as a member of the Air Force Scientific Advisory Board, and an advisor to Air Force studies since 1952.

Licklider became interested in computers after his research in psychoacoustics led him to see the need for a new kind of use for computers and new kinds of man-computer interaction: to, in effect, make "the human and the computer partners in creative thinking."[19] A 1960 paper entitled "Man-Computer Symbiosis" stood as a summary of his vision for the future of computing, developed through his personal research and also influenced by the "tremendous intellectual ferment in Cambridge after World War II."[20] Licklider participated in a circle of Cambridge intellectuals and scientists who came together in weekly meetings led by Norbert Wiener, MIT mathematician and author of *Cybernetics*. Licklider's "Man-Computer Symbiosis," according to historian Paul Edwards, "became the universally cited founding articulation of the movement to establish a time-sharing, interactive computing regime."[21] Bob Taylor, in his preface to a 1990 reprint of "Man-Computer Symbiosis," presented the staggering influence of Licklider's vision in a simple statement: "All users

---

[17] Fubini was Director of Defense Research and Engineering (DDR&E). The DDR&E was the only step in the chain of command between the director of ARPA and the Secretary of Defense. Jack Ruina was ARPA director between February 1961 and September 1963.

[18] Jack Ruina interview, April 20, 1989, in Cambridge, MA, conducted by William Aspray, from transcript. Charles Babbage Institute, University of Minnesota, Minneapolis.

[19] Arthur L. Norberg and Judy E. O'Neill, *A History of the Information Processing Techniques Office of the Defense Advanced Research Projects Agency* (Minneapolis, MN: The Charles Babbage Institute, 1992), 39.

[20] J. C. R. Licklider interview, October 28, 1988, in Cambridge, MA, conducted by William Aspray and Arthur Norberg, from transcript. Charles Babbage Institute, University of Minnesota, Minneapolis.

of interactive computing and every company that employs computer people owe him a great debt."[22]

Historians of IPTO and of the major networking projects it supported (Project MAC and the ARPANET) marvel at the organizational and managerial accomplishments of such a small office.[23] Four features of IPTO's unique approach to managing scientific research have direct roots in the postwar science policy consensus: IPTO's reliance on "centers of excellence"; the "consonance of interest" between military command and control research and interactive computing already underway at places like MIT and CMU; IPTO's ability quickly to fund significant projects; and almost unquestioning support of IPTO from ARPA, the DoD, and Congress.

### 1. Centers of Excellence

Since World War II, the concentration of federal support for scientific research in several "centers of excellence" reflected the influence of Vannevar Bush's vision for the distribution of federal support: Bush favored the development of science, by scientists, for the sake of science.[24] According to Robert Fano, a colleague of Licklider's at MIT

[21] Paul N. Edwards, *The Closed World: Computers and the Politics of Discourse in Cold War America* (Cambridge, MA: The MIT Press, 1997), 266.

[22] Preface by Robert W. Taylor "In Memoriam: J.C.R. Licklider 1915-1990," Digital Research Center Research Report #61, Palo Alto, CA, August 7, 1990. In an interview with William Aspray, Taylor said, "I think most of the significant advances in computer technology, especially in the systems part of computer science over the years… were simply extrapolations of Licklider's vision. They were not really new visions of their own. So he's really the father of it all." Robert Taylor interview, February 28, 1989, in Palo Alto, California, conducted by William Aspray, from transcript. Charles Babbage Institute, University of Minnesota, Minneapolis.

[23] Admirers of the ARPA and IPTO approach include Barber Associates, *ARPA*, Norberg and O'Neill, *Transforming Computer Technology*, and Michael Hauben and Ronda Hauben, *Netizens: On the History and Impact of the Usenet and the Internet* (Los Alamitos, CA: IEEE Computer Society Press, 1997). IPTO's 1963 budget of about 12 million dollars was less than 5% of ARPA's total budget (see Norberg and O'Neill, *Transforming Computer Technology*, 22).

[24] Kevles, "The National Science Foundation," and Harvey Brooks, "Evolution of the U. S. Science Policy Debate: From Endless Frontier to the Endless Resource," unpublished paper presented at Columbia University Seminar on Science Policy, December 9, 1994.

and a recipient of IPTO funding as director of MIT's Project MAC, Licklider believed

very deeply in this concept:

> Licklider's idea was to create what he called a Center of Excellence… You
> know, the place where you breed very good genius… This notion of
> several centers of excellence around the United States was approved at the
> very top [of DoD]… I know that Gene Fubini… was very much behind it.
> So, of course, was Ruina.[25]

This strategy already had created massive Cold War research facilities at American universities, most spectacularly at MIT and Stanford, but also at Berkeley, Michigan,

and Caltech.[26]

Licklider, when asked to name his major accomplishments at IPTO between 1962

and 1964, responded: "I think that I found a lot of bright people and got them working

in this area, well enough to almost define this area. I got it moving… It was more than

just a collection of bright people working in the field. It was a thing that organized

itself a little bit into a community," a community that by the 1990s would be known as

the pioneers of the Internet. As Taylor recognized in a 1990 tribute to Licklider: "The

leaders he chose twenty-five years ago now read like a Who's Who of computing

research."[27]

### 2. "A great consonance of interest'

Licklider's explanation of his intellectual interest in directing IPTO made it very

clear that he viewed the opportunity to work with the military as a chance to advance

---

[25] Robert M. Fano interview, April 20-21, 1989, in Cambridge, MA, conducted by Arthur L. Norberg, from
transcript. Charles Babbage Institute, University of Minnesota, Minneapolis.
[26] Stuart W. Leslie, *The Cold War and American Science: The Military-Industrial-Academic Complex at MIT and
Stanford* (New York: Columbia University Press, 1993). It is worth noting that the first four directors of IPTO, in
office from 1962-1975, were all affiliated with the MIT Lincoln Lab prior to their arrival at ARPA.
[27] Taylor, "In Memoriam."

his evangelical mission, his "religious conversion to interactive computing."[28]  His

recollections in 1989 reflect a singular anticipation of the future of  interactive

computing.  "Every time I had the chance to talk," he later recalled of his first days at

ARPA, "I said that the mission is interactive computing."  Licklider responded to a

question asking why he pushed it so strongly by saying, "I was just a true believer. I

thought, this is going to revolutionize how people think, how things are done."[29]  This was as true for his academic colleagues as it was for the military.  Charles Herzfeld,

ARPA administrator from 1961-1966 and Director from June 1965 to March 1967,

remembered Licklider giving a "series of introductory lectures to modern computing"

beginning in 1962.  Herzfeld recalled that the lectures spoke to the broader values of

making "computing accessible, to make it more efficient, to really use the power of

computers that were available."[30]

Licklider's vision meshed well with ARPA Director Jack Ruina's understanding

of the potential for new forms of computing.  Ruina recalled,

> From my experience in the Pentagon of a couple of years, what I found
> was that the growth of computer technology, hardware technology,
> clearly was exceeding what people knew what to do with it… To my
> mind, the issue at that time was how to explore the potential power that
> was growing in hardware for applications other than straight number
> crunching.[31]

In his discussions with Fubini and Ruina before he moved to ARPA, Licklider

confirmed Ruina's experience:

---

[28] Howard Rheingold, *Tools for Thought*, (Cambridge, MA: The MIT Press, 1985, reprint 2000), 138.  The
evocative language of "religious conversion" is also expressed in Licklider interview, Charles Babbage Institute, and
in Lee and Rosin, "Project MAC Interviews."

[29] Licklider interview, Charles Babbage Institute.

[30] Charles M. Herzfeld interview, August 6, 1990, in Washington, D.C., conducted by Arthur L. Norberg, from
transcript. Charles Babbage Institute, University of Minnesota, Minneapolis.

[31] Jack Ruina interview, April 20, 1989, in Cambridge, MA, conducted by William Aspray, from transcript. Charles
Babbage Institute, University of Minnesota, Minneapolis.

the problems of command and control were essentially problems of man computer interaction… Fubini essentially agreed 100% with that and so did Ruina… Why didn't we develop an interactive computing? If the Defense Department's need for that was to provide an underpinning for command and control, fine. But it was probably necessary in intelligence and other parts of the military too. So, we essentially found that there was a great consonance of interest here, despite the fact that we were using different terms we were talking about the same thing.[32]

Licklider's approach to developing computing for both military uses and to advance the general state of the field is clearly expressed in Licklider's 1963 memo to IPTO's researchers, whom he addressed as the "Members and Affiliates of the Intergalactic Computer Network":

The fact is, as I see it, that the military greatly needs solutions to many or most of the problems that will arise if we tried to make good use of the [computing] facilities that are coming into existence. I am hoping that there will be, in our individual efforts, enough evident advantage in cooperative programming and operation to lead us to solve the problems and, thus, to bring into being the technology that the military needs.[33]

In retrospect, Licklider's achievement was to articulate his vision for interactive computing in a way that would appeal to both the military and to academic and private computer researchers. Licklider's work as director of IPTO shows how he was able to create a consensus around his vision to explore new frontiers:

The technical need is definite in command and control, and also urgent; at the same time, in almost every area of human activity that involves information processing, there is the possibility of a profound advance, which will be almost literally an advance in the way of thinking.[34]

---

[32] Licklider interview, Charles Babbage Institute.

[33] Licklider Papers, MIT Archives, Box 7, "Memorandum for Members and Affiliates of the Intergalactic Computer Network."

[34] Licklider, as quoted in Norberg and O'Neill, *History of IPTO*, 39. In addition to aiding in the rapid development of command and control systems and of academic computing research itself, Licklider's work during this time period reflected his humanitarian commitments to using the power of computing to help education, libraries, problem-solving, access to information, and communication among people. See especially Licklider Papers, MIT Archives, Box 7, "Mechanized Information Systems and Their Relation to Education." Lick's extensive work on digital libraries led to the publication of J. C. R. Licklider, *Libraries of the Future* (Cambridge, MA: The MIT Press, 1965). It also influenced later papers, including J.C.R. Licklider and Robert Taylor, "The Computer as a Communication Device," *Science and Technology*, April, 1968, reprinted in "In Memoriam: J. C. R. Licklider, 1915-1990," 21-41.

### 3. "We were light on our feet"

IPTO's third organizational advantage was that it could move quickly to support research, another principle enshrined in the science policy consensus canon in the wake of *Sputnik.* Ivan Sutherland, director of IPTO from June 1964 to June 1966, spoke about the ease, on both an ideological and administrative level, with which ARPA could dispense funding:

> My understanding of the mechanism was that if I could convince the ARPA director that something was sensible to do, it got done… I always felt the principal hurdle was to convince the ARPA director that it was a sensible task. I guess Al Blue [IPTO administrator, 1965-1977] and the administrative machinery ran smoothly enough that the ARPA orders got written.[35]

On his role in the "administrative machinery," Blue recalled fondly, "ARPA was a unique place. I am sure you have heard the fact that we could get an idea in the morning and have the guy working under a Letter of Intent by 4:00 the next day."[36] Charles Herzfeld, director of ARPA from June 1965 to March 1967, said that "ARPA was the only place in town where somebody could come into my office with a good idea and leave with a million dollars at the end of the day."[37] Indeed, this is precisely what happened in 1966 to Bob Taylor, director of IPTO at the time. Taylor wanted to convince Herzfeld of the necessity for creating a network that would connect IPTO computers. Such an idea, directly in line with Licklider's philosophy, would spawn greater connections within the IPTO research community and save the DoD from having to buy more and more expensive mainframes. After a brief discussion, Taylor left Herzfeld's office with a million dollars for the network. It was a stunning achievement, even at ARPA. An account in *Where Wizards Stay Up Late* of "The Fastest Million Dollars" aptly characterizes Taylor's surprise at how easy it was: "'Jesus Christ,' he said to himself softly. 'That only took twenty minutes.'"[38]

---

[35] Ivan Sutherland interview, May 1, 1989, in Pittsburgh, PA, conducted by William Aspray, from transcript. Charles Babbage Institute, University of Minnesota, Minneapolis.
[36] Allan Blue interview, June 12, 1989, in Minnetonka, MN, conducted by William Aspray, from transcript. Charles Babbage Institute, University of Minnesota, Minneapolis.

### 4. "Benign neglect" and "No concern with relevance"

ARPA directors, officers, and staffers universally praised ARPA's practice of delegating authority to program managers, who would in turn allow their contractors to proceed with their research, more or less undisturbed and on their own terms. This phenomenon was amplified within IPTO, which, due to its relatively small size and budget, did not capture the attention of ARPA directors or their superiors in DoD.

Ruina stressed in a 1989 interview that IPTO "was not a major program," and that, as director of ARPA, he was more occupied with ARPA's programs on ballistic missile defense and nuclear test detection. Licklider recalled of Ruina, "[h]e had much bigger fish to fry; this was a small part of his life. I've told him since that I've felt it was a kind of benign neglect." Licklider continued, "I think he [Ruina] also had a feeling, 'I have got 50 or 100 million dollars in getting these nuclear warheads back in; why can't I have 10 or 15 in command and control…"[39] In light of the United States' decisive use of information technologies in the 1991 Gulf War and when considering the impact of the computer networking on the global economy, the decision to put "10 or 15 million" annually into research exploring "interactive computing" was one of the best investment decisions of American history.

---

[37] Herzfeld interview, Charles Babbage Institute.
[38] Katie Hafner and Matthew Lyon, *Where Wizards Stay up Late: The Origins of the Internet* (New York: Touchstone, 1996), 42.

IPTO had the autonomy and the money to fund what they perceived to be smart people and good research. As Allan Blue recalled of the era before the passage of the Mansfield Amendment in 1970,[40] IPTO was

> an environment in which you gave money to good people and expected good results. There was no concern with relevance… if a program manager has a good idea, he has got two people to convince that that is a good idea before the guy goes to work. He has got the director of his office and the director of ARPA, and that is it.[41]

Licklider highlighted this approach by pointing out that "my two immediate superiors in the chain of command, Ruina and Sproull, did me the great favor of listening intently long enough to decide that they were really fundamentally in support of what was going on. After that they spent little time heckling me."[42]

IPTO's autonomy was further reinforced by its relative invisibility to the Congressional oversight committees that approved ARPA's budget (IPTO's budget during the 1960s represented between 5 and 10% of the total ARPA budget).[43] By demonstrating the need in the military for flexible, reliable, and effective computing systems, and by focusing on the relevance of IPTO research to a wide variety of military and scientific needs, ARPA directors defended IPTO's budget from further scrutiny by sticking to the rules of the consensus. In other words, it was bad politics in Cold War America to prevent the DoD, and hence ARPA and IPTO, from doing what they thought was best.

---

[39] Licklider interview, Charles Babbage Institute.
[40] On the Mansfield Amendment, see Norberg and O'Neill, *Transforming Computer Technology*, 36-38, and James L. Penick, Jr., Carroll W. Pursell, Jr., Morgan B. Sherwood, and Donald C. Swain, eds., *The Politics of American Science, 1939 to the Present* (Cambridge, MA: The MIT Press, 1973), 338-349.
[41] Blue interview, Charles Babbage Institute.
[42] J. C. R. Licklider, "The Early Years: Founding IPTO," in Thomas C. Bartee, ed., *Expert Systems and Artificial Intelligence: Applications and Management* (Indianapolis: Howard W. Sams & Company, 1988), 222. Robert Sproull succeeded Jack Ruina as ARPA director in 1963.
[43] Norberg and O'Neill, *Transforming Computer Technology*, 22.

Without exception, the individuals involved in these "early days" of networking express pride that their research was supported and allowed to blossom. As Herzfeld reflected in 1990, Licklider

> predicted the future of computing in America remarkably well…
> [Licklider] said, 'We clearly can do the following. It makes sense and we ought to do it, so let's go do it.' And indeed, it happened. Networking, interactive graphics, time-sharing, and all these things that are now so commonplace were in the air, and he saw to it that they would happen.[44]

**Leveraging the Consensus: Taylor, Roberts, and the ARPANET**

ARPA's use of its influence to leverage personnel and projects between 1966 and 1969 illuminates the ways that ARPA used its position within the science policy consensus. Robert Taylor, IPTO Director from June 1966 to March 1969, joined IPTO because he "became heartily subscribed to the Licklider vision of interactive computing."[45] Taylor, convinced that Larry Roberts in the MIT Lincoln Lab would be the ideal candidate to direct a proposed IPTO computer network, found no success in his initial attempts to bring Roberts to IPTO. Once Taylor realized in "September or October of 1966" that the DoD provided 51% of the Lincoln Lab's funding, he asked Herzfeld (ARPA Director at the time) to call the Lincoln Lab's director and "tell him it's in Lincoln Lab's and ARPA's best interests for him to tell Larry Roberts to come down and do this."[46] Although he was "not particularly eager for the job," Roberts was working in the Pentagon for IPTO by January 1967.[47]

---

[44] Herzfeld interview, Charles Babbage Institute.
[45] Taylor interview, Charles Babbage Institute.
[46] Taylor interview, Charles Babbage Institute.
[47] L. Roberts, "Expanding AI Research and Founding ARPANET," in Bartee, ed., *Expert Systems and Artificial Intelligence*, 229.

What was initially a move that Taylor called "blackmail" soon became an opportunity for Roberts, 28 years old at the time, to pursue his goal of sharing the research of the computer science community with a broader audience. Roberts recalled,

> I was also coming to the point of view, separately from that [Herzfeld's "51%" phone call], that this research was not getting to the rest of the world… I was feeling we were now probably twenty years ahead of what anybody was going to use and still there was no path for them to pick up… So I was really feeling a pull towards getting more into the real world, rather than remain in that sort of an ivory tower… So they eventually convinced me it was a good idea—particularly after Lincoln said that was the best place for me.[48]

Some of the tensions as well as the benefits of ARPA's leverage are evident within Roberts' statement. Roberts' recollections make it clear that ARPA—specifically IPTO—had by 1966 established itself as the most advanced and influential sponsor and organizer of computer research, and that its position left him little choice—both professionally and intellectually—other than moving to the Pentagon.

Roberts quickly learned to use ARPA's leverage to further the computer networking research agenda he pursued at Lincoln Lab. The objective for the ARPANET, according to Taylor, was to "allow people who were separated geographically to discover and then exploit common interests."[49] A network, which would allow this sort of sharing of IPTO-funded research, would also enable the sharing of computing power as well as "to get these timesharing communities—which had been built up locally as a community… to build metacommunities out of these by connecting them."[50] The reflections of Taylor and Roberts indicate that the motivating values behind the creation of the ARPANET were to share IPTO research, IPTO resources, and to broaden the computer research community.

---

[48] Roberts interview, Charles Babbage Institute.
[49] Robert Taylor, quoted in Stephen Segaller, *Nerds 2.0.1: A Brief History of the Internet* (New York: TV Books, 1998), 58.
[50] Taylor interview, Charles Babbage Institute.

The "resource-sharing" element of the ARPANET is another example of IPTO's leveraging of their power within the Cold War science policy consensus. Upon presenting the plans for the ARPANET to IPTO Principal Investigators (PIs) and the requirement that all IPTO-funded sites must participate, Roberts encountered "tremendous resistance" from the PIs.[51] Despite the support of Herzfeld and Licklider, Taylor remembered that "most of the people I talked to were not initially enamored with the idea."[52] Roberts, the object of ARPA's leverage only months earlier, skillfully turned that leverage to his advantage:

> Well, the universities were being funded by us, and we said, "We are going to build a network and you are going to participate in it. And you are going to connect it to your machines. By virtue of that we are going to reduce our computing demands on the office. So that you understand, we are not going to buy you new computers until you have used up all of the resources of the network. So over time we started forcing them to be involved, because the universities in general did not want to share their computers with anybody.[53]

Just as Roberts and his boss at Lincoln Lab in 1966 had no choice but to give in to ARPA's demands, the IPTO-funded PIs were more or less coerced into making their computers part of the ARPANET. But, also like Roberts, the PIs came to appreciate the results of the "arm-twisting." According to Roberts, the pioneers of Artificial Intelligence at MIT and Stanford, Marvin Minsky and John McCarthy, initially

> hated the fact that somebody else might use their computer. They found that to be a gross invasion of their privacy. Eventually, they applauded it because they could copy each others' papers! They could get at the

---

writing of each others' groups and that sort of thing, so they found the communication valuable to their own groups. But initially, they were very negative. I still remember them fighting all along the way.[54]

Roberts' management of the ARPANET would immediately be characterized by what Janet Abbate identified as "informal mechanisms aimed at creating and reinforcing common values and goals."[55] "The PIs were the primary group we worked with," Roberts recalled, "both in the PI meetings and individually, in order to get a consensus in the community of what was good and bad… pretty much everyone helped in the process, even if it was not conscious to them."[56] The influence of consensus-based development in the history of the Internet between 1969 and the present is a subject outside the scope of this paper that deserves further study. At first glance, however, the use of voluntary consensus standards to create the architecture of the Internet, and the Internet Engineering Task Force's motto, "rough consensus and running code," both testify to the enduring power of "consensus" as a governing value of cyberspace.[57]

## Conclusion

Recent historiography of the Cold War and of Cold War computing points to "containment" metaphors and "closed world discourse" as dominant themes of the period between 1945 and the present. In contrast to these metaphors and themes, the evidence presented in this paper suggests that the theme of "consensus" served as a major factor in the mobilization of science for the national defense, in the creation of ARPA, in the operating philosophy of IPTO, in the hiring of scientists such as Licklider and Roberts, and in the intellectual formation and development of the ARPANET. A greater awareness of the ways that "consensus" worked in this period—the "pre-history" of the Internet—provides a richer context for evaluating the unique features of the Internet, such as its open architecture, collegial culture, and standards-based governance. If we interpret "closed world discourse" as dominant among the values of the Internet's design, such features as the Internet's open and decentralized architecture would be more difficult to explain.

---

[54] Roberts interview, Charles Babbage Institute.
[55] Janet Abbate, *Inventing the Internet* (Cambridge: The MIT Press, 1999), 69.
[56] Roberts interview, Charles Babbage Institute.
[57] See for example Carl F. Cargill, *Open Systems Standardiztion: A Business Approach* (Upper Saddle River, NJ: Paladin Consulting, Inc., 1997).

Attention to "consensus" as another founding principle of the ARPANET's design helps us to account for its open and decentralized architecture as well as its consensus-based governance. It can also be seen as a set of attitudes that created space for human values to be incorporated into the ARPANET's design. Abbate's *Inventing the Internet* highlights several of these values: responsiveness, reliability, expandability, standard protocols to promote interoperability, decentralized complexity through layering, and informal and inclusive management.

Calls for historians to locate the history of the Internet "within its social, political, and cultural contexts"[58] underscore the need for both academics and policymakers—especially those working closely with technical and/or administrative aspects of the Internet—to think broadly when considering the values that have influenced and will continue to influence the design of cyberspace. Between 1957 and 1969, the Cold War consensus about the roles of science and government shaped the creation of the IPTO and the ARPANET. Some of the broader issues of this Cold War consensus included the convictions that the government and the universities should cooperate to advance science as a path toward national progress and in order to protect national security.

---

[58] Roy Rosenzweig, "Wizards, Bureaucrats, Warriors, and Hackers: Writing the History of the Internet," *American Historical Review* 103 (December 1998): 1530-52, and John S. Quarterman, "Revisionist Internet History," *Matrix News*, 9 (4), April 1999, http://www.mids.org/mn/904/large.html (visited September 18, 2001).

In 2001 we may look to the influence of similarly broad issues such as the features of "globalization" that tend toward global economic and social integration and stratification, and frequently pit corporate interests against scientific or "public" interests. The broader themes of globalization already have influenced public policy debates in broadband and wireless deployment and the "digital divide." Security researchers complain of the "chilling effects" of the Digital Millennium Copyright Act on their efforts to create more secure systems and encryption techniques.[59] Scholars of globalization also detect familiar themes in debates about the nature of intellectual property and whether software engineering should follow a proprietary model or, instead, a competing "open source" model.[60] A startling transpartisan alliance between House Majority Leader Dick Armey (R-Texas) and the American Civil Liberties Union highlights the broad-based concern with the increasing threats to privacy posed by public surveillance systems.[61] As historical analysis clearly demonstrates, these and other broad value conflicts will continue to exert significant influence on the design and governance of cyberspace.

---

[59] See the "Chilling Effects Clearinghouse," http://eon.law.harvard.edu/chill (visited September 18, 2001) and the Electronic Frontier Foundation's coverage of Felten vs. RIAA, http://www.eff.org/Legal/Cases/Felten_v_RIAA (visited September 18, 2001).

[60] Lawrence Lessig, *Code and Other Laws of Cyberspace* (New York: Basic Books, 1999) and Eric Raymond, *The Cathedral and the Bazaar: Musings on Linux and Open Source By An Accidental Revolutionary* (Sebastopol, CA: O'Reilly & Associates, Inc., 1999).

[61] See "Proliferation of Surveillance Devices Threatens Privacy: Joint Statement of House Majority Leader Dick Armey, R-TX, And The American Civil Liberties Union," http://www.aclu.org/news/2001/n071101a.html (visited September 18, 2001).